\documentclass[doublespacing]{elsart}

\usepackage{amssymb}
\usepackage{epsfig}
\usepackage{graphicx}

\begin{document}

\begin{frontmatter}



\title{Hybrid systems with virtual cathode for high power microwaves generation}


\author{V.G. Baryshevsky,}
\author{A.A. Gurinovich\corauthref{cor1}}
\corauth[cor1]{gur@inp.minsk.by; sa\_shen\_ka@yahoo.com}

\address{Research Institute for Nuclear Problems,
Belarus State University, 11 Bobruyskaya Str., Minsk 220050,
Belarus}

\begin{abstract}
It is shown that use of a photonic crystal enables to construct
several types of hybrid systems with virtual cathode, which could
radiate due to different radiation mechanisms (bremsstrahlung and
diffraction (transition) radiation) with different frequencies.
Also mentioned that photonic crystal makes possible to create
phase-locked array of generators.
\end{abstract}


\end{frontmatter}



Interest to high power microwave (HPM) sources has emerged in
recent years due to revealing new applications and offering novel
approaches to existing applications.

Vacuum electronic sources, which convert the kinetic energy from
an electron beam into electromagnetic field energy, are a natural
choice for generating HPM.

The high current density electron beam, once generated, propagates
through an interaction region, which converts the beam's kinetic
energy to HPM.
It is the particular nature of the interaction that distinguishes
the various classes of sources.

 High power HPM
sources generating high electromagnetic power density require the
high power densities in the electron beams, where space-charge
effects are essential.

When the magnitude of the current $I_b$ of an electron beam
injected into a drift tube exceeds the space-charge-limiting
current $I_{limit}$, an oscillating virtual cathode (VC) is formed
\cite{RuhadzeUFN,Selemir.rev1}.

According to \cite{RuhadzeUFN} the following formula gives a good
approximation for the space-charge-limiting current:
\begin{equation}\label{I_limit}
    I_{limit}(kA)=\frac{m c^3}{e} G (\gamma^{2/3}-1)^{3/2}
\end{equation}
 where $m$ and $e$ are the mass and charge of an electron, $c$ is the speed of light, $\gamma$ is the Lorentz factor of the electron beam
 and $G$ depends on the geometry \cite{RuhadzeUFN,NeskolkoVC}.
For example, for an annular electron beam in a cylindrical drift
tube $G$ reads as follows \cite{NeskolkoVC}:
\begin{equation}\label{G}
    G=\frac{1}{(\frac{r_b-r_b^{in}}{r_b}+2 \ln \frac{R}{r_b}) (1- sech \frac{\mu_1 L}{2
    R})},
\end{equation}
where $r_b$ and $r_b^{in}$ are the outer and inner radius of the
electron beam, $R$ and $L$ are the radius and length of the
cylindrical drift tube and $\mu_1$ is the first root of the Bessel
function $J_0 (\mu)=0$.

\begin{figure}[h]
\epsfxsize = 10 cm \centerline{\epsfbox{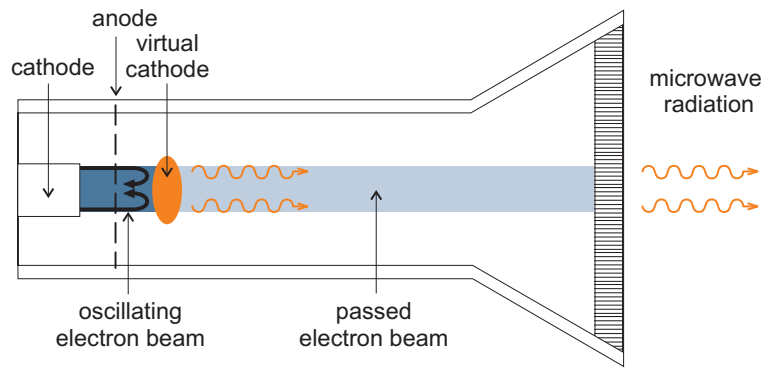}} \caption{An
oscillator with virtual cathode} \label{fig1}
\end{figure}

When oscillating virtual cathode is formed two types of electrons
exist: those oscillating in the vircator potential well and
passing through the vircator area (see Fig.\ref{fig1}).

For electrons oscillating in the area ''cathode-anode-virtual
cathode'' two radiation mechanisms provide radio-frequency signal
\cite{Giri}:

1. one radiation mechanism originates from the oscillations of the
reflected electrons about the anode foil (electron oscillations in
the potential well ''cathode-anode-virtual cathode'').
A microwave signal is generated at  a frequency of roughly
$\frac{c}{2 d_d}$, where $d_d$ is the anode-to-cathode spacing.

2. The other radiation mechanism is the oscillation of the virtual
cathode at a frequency near the plasma frequency $\omega_p$ of the
space charge density that is formed. That is
\begin{equation}
\omega_p=\sqrt {\frac{4 \pi n_e e^2}{m}}
 \label{omegap}
\end{equation}
where $n_e$ is the number density of the electrons in the space
charge configuration (in the plane of the
anode grid) \cite{Giri}.

The essence of the above radiation mechanisms is bremsstrahlung
radiation ensuing from electron deceleration.

In the present paper we emphasize that bremsstrahlung radiation
from electrons oscillating in an electron beam with a virtual
cathode is accompanied by transition radiation, which origin is
electron velocity rather then acceleration. Use of a photonic
crystal enables to construct several types of hybrid systems with
virtual cathode, which could radiate due to different radiation
mechanisms (bremsstrahlung and diffraction (transition) radiation)
with different frequencies.

In vircator systems grid cathode and anode (or anodes) are
commonly used
\cite{NeskolkoVC,Selemir_patent,87Selemir,Selemir_patent2}.
 Electron beam oscillates making electrons periodically crossing the grid
 anodes and cathode (see, for example \cite{87Selemir}).
It is transition radiation that occur when electrons pass through
a border between two media with different indices of refraction.
It is worth noting that periodical excitement of transition
radiation from electrons oscillating in vircator is similar to
diffraction radiation from a charged particle in a periodic
structure.
As a result in a system with oscillating virtual cathode the
vircator radiation, which actually is electron beam
bremsstrahlung, is accompanied by radiation excited by additional
mechanism due to transition (diffraction) radiation from
oscillating current passing through the grid anodes (cathode).

Let us turn to that part of the beam, which passes through the
virtual cathode area.
Recall that the oscillation of the virtual cathode can produce a
highly modulated electron beam, and, as a result, the energy from
the bunched transmitted beam can be recovered using slow-wave
structures \cite{SelemirTWT}.

\begin{figure}[h]
\epsfxsize = 10 cm \centerline{\epsfbox{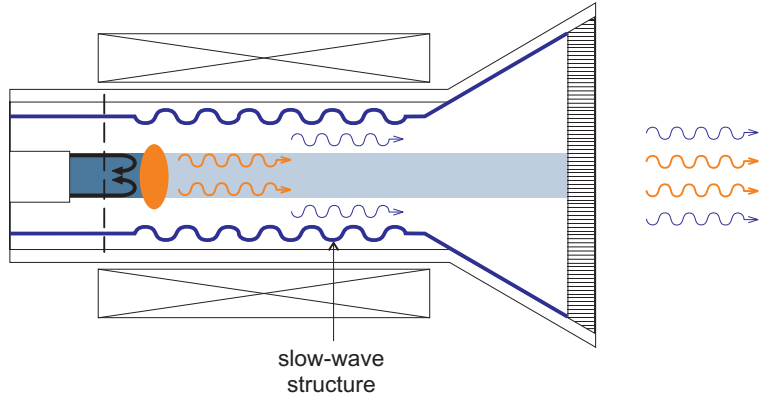}} \caption{Hybrid
system ''vircator + travelling wave tube'' \cite{SelemirTWT}}
\label{fig2}
\end{figure}

Interaction of the electron beam with the slow-wave structure in,
for instance, conventional TWT pose special challenges:
interaction is sufficiently effective only for electrons moving at
the distance $\delta$ from the slow-wave structure surface
\begin{equation}\label{delta}
\delta \le \frac{{\lambda \beta \gamma} }{{4\pi} },
\end{equation}
$\delta $ is the so-called  beam impact parameter, $\lambda $ is
the radiation wavelength, $\beta = v /c$ , $v$ is the electron
beam velocity, $\gamma $ is the electron Lorentz-factor.
For example, for electrons with the energy 250 keV ($\beta=0.74$
and $\gamma=1.49$) and radiation wavelength $\lambda= 10 mm $
(frequency 30 GHz) the impact parameter $\delta \approx 0.9 mm$.
It means that for efficient radiation generation an annular
electron beam with the thickness $\Delta \leq \delta$ otherwise
only part of the beam would contribute to radiation.

As we have shown in
\cite{VFELreview,grid-t,grid-ex,grid-NIM,new-grid-experiment,grid-FEL06-ex,grid-FEL06-th,grid-FEL07-ex}
 this challenge
can be overcome applying a photonic crystal, formed by metallic
grids (grid travelling wave tube(grid TWT), grid volume free
electron laser(grid VFEL)).

What is more, in accordance with
\cite{27Selemir,28Selemir,29Selemir} application of metallic
inserts (meshes, grids and so on) inside a resonator enables
increasing of the electron beam limit current.

Therefore, in the grid TWT (grid VFEL) presence of the metallic
grid (photonic crystal) serves both for forming of resonator,
where interaction of the beam and radiation occurs, and for
potential balancing that makes possible to increase the beam
vacuum limit current.

And for the grid TWT (grid VFEL) with the supercritical current
the electron beam executes compound motion exciting two radiation
mechanisms contributing to radiation: bremsstrahlung of
oscillating electrons and diffraction (transition) radiation from
downstream electrons interacting with the periodic grid
structure(photonic crystal).

This means that the hybrid system "vircator + grid TWT (grid
VFEL)" arise by analogy with \cite{SelemirTWT}, where several
vircators could appear due to presence of several anode grids (see
also \cite{NeskolkoVC}).
But in contrast to the system \cite{NeskolkoVC} the hybrid system
"vircator + grid TWT (grid VFEL)" uses periodically placed grids
with either constant \cite{grid-ex,grid-FEL06-ex,grid-FEL06-th} or
variable period \cite{grid-FEL07-ex}.

Frequency of diffraction radiation excited by an electron beam in
a periodic structure with the period $d$ is determined by the
condition
\begin{equation}
\omega - \vec{k} n({k}){\vec{v}}= \vec{\tau} \vec{v} ~,
\label{eq:3-2}
\end{equation}
where $\vec{v}$ is the electron beam velocity, $\vec{\tau}$ is the
reciprocal lattice vector ($|\vec{\tau}|=\frac{2 \pi p}{d}$),
$n(k)$ is the refraction index of periodic structure, $p$ is an
integer number ($p=1,2,3,...$).

When the electron beam velocity $\vec{v}$ is parallel to the
reciprocal lattice vector $\vec{\tau}$ (\ref{eq:3-2}) reads
\cite{grid-NIM}
\begin{equation}\label{frequency}
    \omega=\frac{2 \pi p \cdot v}{d(1-\beta n(\omega,k) \cos \theta)}
\end{equation}

\begin{figure}[h]
\epsfxsize = 15 cm \centerline{\epsfbox{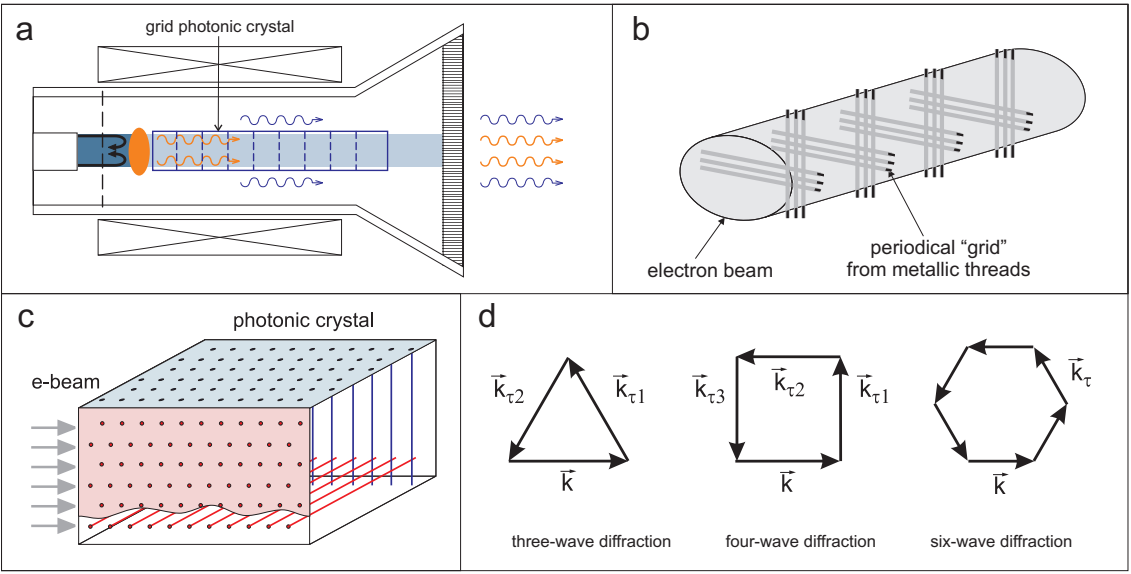}} \caption{Grid
TWT (grid VFEL) and photonic crystal arrangement} \label{fig3}
\end{figure}

Essentially, photonic crystal in the grid TWT (grid VFEL) is
transparent for radiation as well as for electron beam (see
Fig.\ref{fig3}).
Moreover, several diffracted waves could exist in photonic crystal
(see Fig.\ref{fig3}d) that makes possible to introduce feedback in
such a system at the frequency of diffraction radiation and,
hence, to couple several hybrid ''vircator + grid TWT (VFEL)''
generators making a phase-locked source, in which diffracted waves
from one photonic crystal (grid resonator) excites oscillations in
the neighbor resonators.

\begin{figure}[h]
\epsfxsize = 10 cm \centerline{\epsfbox{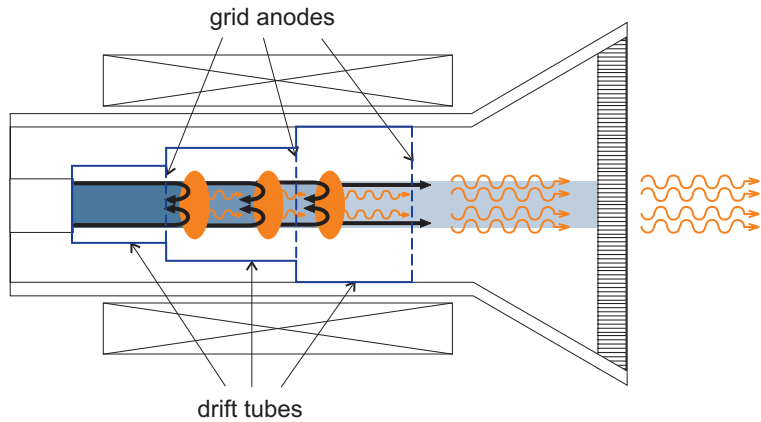}}
\caption{Several vircators \cite{NeskolkoVC}} \label{fig4}
\end{figure}

The proposed grid systems drastically differ from the system
\cite{NeskolkoVC}, where several grids serve only for forming of
several vircators (see Fig.\ref{fig4}).

Of course, radiation from hybrid generator ''vircator + grid TWT
(VFEL)'' can be excited by several electron beams similar
phase-locked array.

\begin{figure}[h]
\epsfxsize = 10 cm \centerline{\epsfbox{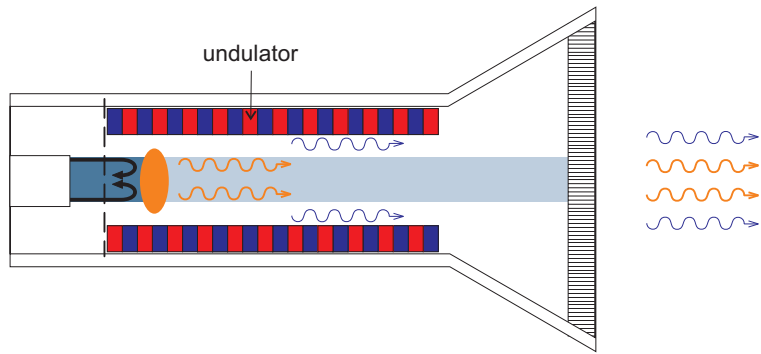}} \caption{Hybrid
system ''vircator + undulator FEL''} \label{fig5}
\end{figure}

The bunched electron beam passed through the virtual cathode area
can be also used for excitation of free electron laser (ubitron)
(see Fig.\ref{fig5}) oscillation contributing to the radiation
power.

\section{Conclusion}

It is shown that use of a photonic crystal enables to construct
several types of hybrid systems with virtual cathode, which could
radiate due to different radiation mechanisms (bremsstrahlung and
diffraction (transition) radiation) with different frequencies.
Also mentioned that photonic crystal makes possible to create
phase-locked array of generators.

\end{document}